%% file: main.tex
\documentclass{extarticle}
\usepackage{spconf,amsmath,graphicx}
\usepackage{algpseudocode}
\usepackage{algorithm}
\usepackage{appendix}
\usepackage{url}
\usepackage{multicol}

\usepackage{multirow}
\usepackage{color}

\def\proposed{{\mathcal PRS}}

    \title{Partial Rank Similarity Minimization Method for Quality MOS Prediction of Unseen Speech Synthesis Systems in  Zero-Shot and Semi-supervised setting}
\name{Hemant Yadav\sthanks{The work was performed while at National Institute of Informatics (NII), Tokyo, Japan.}, Erica Cooper, Junichi Yamagishi, Sunayana Sitaram, Rajiv Ratn Shah}
\address{\{hemantya,rajivratn\}@iiitd.ac.in, \{ecooper,jyamagis\}@nii.ac.jp, sunayana.sitaram@microsoft.com}

\copyrightnotice{979-8-3503-0689-7/23/\$31.00~\copyright2023 IEEE}
\begin{document}
\maketitle
\begin{abstract}

This paper introduces a novel objective function for quality mean opinion score (MOS) prediction of unseen speech synthesis systems. The proposed function measures the similarity of relative positions of predicted MOS values, in a mini-batch, rather than the actual MOS values. That is the partial rank similarity is measured ($\proposed$) rather than the individual MOS values as with the L1 loss. Our experiments on out-of-domain speech synthesis systems demonstrate that the $\proposed$ outperforms L1 loss in zero-shot and semi-supervised settings, exhibiting stronger correlation with ground truth. These findings highlight the importance of considering rank order, as done by $\proposed$, when training MOS prediction models. We also argue that mean squared error and linear correlation coefficient metrics may be unreliable for evaluating MOS prediction models.
In conclusion, $\proposed$-trained models provide a robust framework for evaluating speech quality and offer insights for developing high-quality speech synthesis systems. Code and models are available at 
\url{github.com/nii-yamagishilab/partial_rank_similarity/}

\end{abstract}

\section{Introduction}
\label{sec:intro}
\input{sections/introduction}

\section{Methodology}
\label{sec:method}
\input{sections/method}

\section{Experiments}
\label{sec:experimentation}
\input{sections/experimentationdetails}

\label{sec:results}
\input{sections/results}

\section{Discussion}
\input{sections/discussion}

\section{Conclusions and Future Work}
\label{sec:conclusions}

This paper introduced the $\proposed$ method for predicting Mean Opinion Score (MOS) values given an audio sample. By considering the relative position in the ranking of MOS values within each training batch, $\proposed$ provides a novel approach to capture ranking information. In this study, we also present E-$\proposed$, an extension of $\proposed$ to incorporate samples for comparison beyond the current batch size for better generalization. Comparative evaluations with existing methods demonstrated comparable performance on in-domain and superior performance on OOD datasets. The experimental results highlight the generalization ability of $\proposed$ in MOS prediction tasks. Contrary to popular belief, we posit that MSE and LCC are unreliable evaluation metrics for comparing MOS prediction systems. We also demonstrated that performance can be further improved if a better selection method is used in the semi-supervised finetuning stage, similar to \cite{de2016highconfidence}. 

Our future work includes the following ideas. Instead of averaging the features from the last layer of Wav2Vec2.0, using a recurrent neural network (RNN) as the last layer during finetuning as proposed by \cite{saeki22c_interspeech} may also improve the performance. We will also consider investigating the use of attention to average the frame-level features. Furthermore, similar to \cite{tseng22b_interspeech}, additional unsupervised domain-adaptive pre-training of the Wav2Vec2.0 model to learn better features may result in performance improvements in the zero-shot and few-shot settings. Lastly, a better selection algorithm to sample the pseudo MOS values in the semi-supervised setting on the basis of some heuristics could lead to improvements as well. We also intend to explore whether employing an ensemble of MOS models enhances the reliability of predictions, resembling the MOS test conducted with multiple human annotators.

\section{Acknowledgments}
The authors thank Xin Wang
for his valuable feedback, and Aidan Pine for making his listening test data available for this work. This study is supported by JST CREST Grant Number JPMJCR18A6 and by MEXT KAKENHI grant 21K11951. RR Shah is partly supported by the CAI and CDNM at IIIT Delhi, India. Hemant Yadav is supported by Microsoft Research India PhD Fellowship program.

\bibliographystyle{IEEEbib}
\bibliography{IEEE}

\onecolumn
\clearpage
\appendix
\section*{\parbox{\linewidth}{\centering APPENDIX}}
\def\thesection{\Alph{section}}
\input{sections/appendix.tex}

\end{document}

%% file: sections/introduction.tex
Recent advances in machine learning have significantly improved synthesized speech, which consequently has become more integrated into our daily lives. Unlike machine translation, which uses BLEU score \cite{papineni-etal-2002-bleu} for algorithmic evaluation, text-to-speech (TTS) synthesis and voice conversion (VC) heavily rely on human ratings from listening tests. Crowdsourcing \cite{eskenazietal2013} and web-based tests have expanded participant pools and accelerated experimentation; however, these are still more costly and time-consuming than automated evaluation metrics. Thus, there is increasing interest in developing reliable objective quality measures for synthesized speech.

\begin{figure}[!t]
\centering
\scalebox{0.97}{
\includegraphics[trim=4cm 2cm 3cm 1cm, clip,width=1\columnwidth]{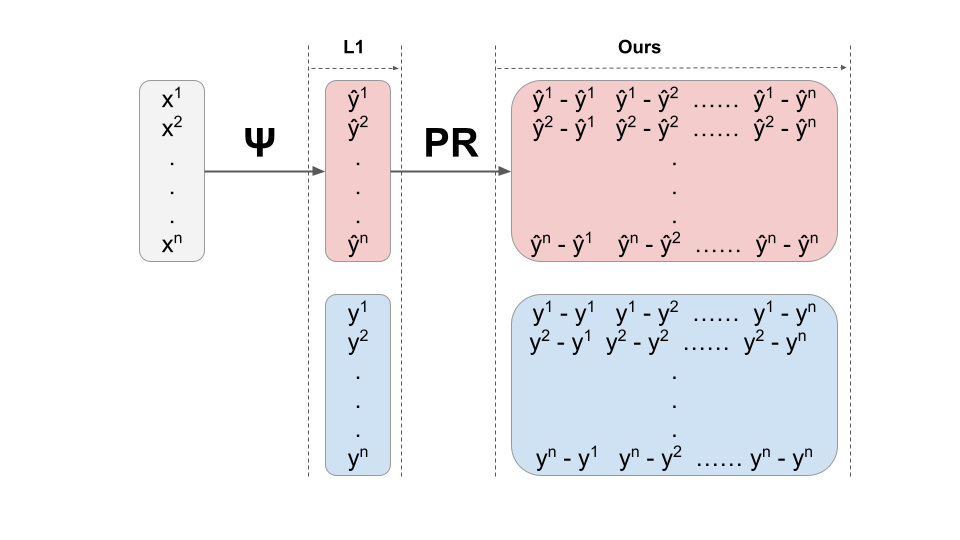}}
\caption{A typical MOS prediction pipeline. It consists of a function approximator $\psi$ to predict the MOS scores, given an audio file. Standard practice is to calculate the L1 loss using the predictions and ground-truth MOS values. In this work, we apply a partial ranking function $PR$ and then apply the p-norm loss over the output matrices of prediction and ground-truth MOS values. 
}
\label{fig: Mos prediciton pipeline.}
\end{figure}

Mean opinion score (MOS) serves as an attractive evaluation methodology for researchers due to its ability to provide a single, easily comparable numerical result. In a MOS test, listeners evaluate synthesized samples one by one and assign them an integer rating on a scale (e.g., 1-5) on the basis of some criteria such as naturalness. All ratings per system are averaged together to obtain a final mean score. With the recent advances in machine learning, attention has turned to data-driven synthesized speech quality prediction -- in particular, automatic MOS prediction. Early works on neural network-based data-driven MOS prediction \cite{mosnet,leng2021mbnet,williams2020comparison} found that although MOS ratings from the same listening test as the training data could be well-predicted, these models do not generalize well to data from other listening tests due to differences in the listener pool, testing interface, systems under consideration, and many other factors outlined by \cite{zielinski2008some}. The authors of \cite{baseline, huang22voicemos} showed that finetuning self-supervised learning (SSL) based models for speech, such as Wav2Vec2 \cite{wav2vec2},  could increase the generalization ability of automatic MOS predictors on out-of-domain (OOD) datasets. To mitigate the domain mismatch between pretrained SSL models, which have only seen examples of natural speech, and the MOS prediction task for synthesized speech, \cite{tseng22b_interspeech} conducted domain-adaptive pretraining  \cite{gururangan-etal-2020-dont}. They show improvements on an OOD dataset, most notably in the zero-shot and few-shot settings. However, predicting unseen systems from OOD listening tests remains challenging. In fact, this is a crucial scenario for researchers and engineers utilizing automatic quality predictors. They often develop and assess new, unseen systems, including those for different languages, including low-resource languages.

It was noted in \cite{baseline} that in the zero-shot prediction scenario, where the model has not been finetuned on any labeled data from the target listening test, that mean squared error (MSE) can be very high even when the correlations with true MOS values are reasonable. We hypothesize that, indeed, predicting the correct {\em ordering} of synthesis systems with respect to their naturalness is more meaningful than predicting the absolute MOS values. As an example, if we use a rating scale from {1-5} and keep the rank order of MOS ratings for the audio samples the same but shift and skew the overall distribution of their scores towards either end of the scale to simulate listener and other contextual biases, then MSE will increase substantially, although ranking-based correlations will remain high. 
Using metrics such as MSE and linear correlation coefficient (LCC), which are dependent on the absolute MOS values, can be misleading in evaluating different MOS predictors, especially in the zero-shot OOD setting. 

In the same spirit, the authors of UTMOS \cite{saeki22c_interspeech} proposed a loss that enforces correct rank order, obtaining conclusive improvement on an OOD dataset and supporting our hypothesis. However, the authors of UTMOS in their paper did not discuss why the rank order is important nor did they investigate the performance of their loss function in zero-shot or semi-supervised settings. In contrast, we justify our loss function using the partial rank order within a mini-batch and show that MSE and LCC are unreliable metrics for evaluating MOS prediction systems. The core idea of our method is most similar to UTMOS \cite{saeki22c_interspeech}. The most notable difference is in the loss formulation. Their loss contains a margin term to avoid penalizing small errors, but which has the consequence that the loss could be zero even if the rank order is incorrect. This is an undesirable behavior when predicting MOS values. Lastly, different from prior work, we also study the effect of extending the total number of comparisons beyond the current batch size. The differences between UTMOS and the proposed method will be described in more detail in Section \ref{sec:method}.

In this paper, we propose a method that addresses the challenging case of {\em zero-shot and few-shot quality MOS prediction for unseen, OOD speech synthesis systems.} Rather than focusing on absolute measures, 
we aim to measure {\em similarity of partial rank order matrices obtained from MOS values for multiple (but not necessarily all) samples and systems, particularly in terms of naturalness.} 
Our contributions are as follows: 

\begin{enumerate}

    \item We explain why relative position in the rank order is important to consider when solving the MOS prediction task. 
    \item We formulate a loss function on the basis of the relative position in the rank order that covers parts of systems to be evaluated and call it Partial Rank Similarity ($\proposed$) loss.  
    \item We introduce a BAlanced pseudo MOS (BApMOS) selection approach for choosing unlabeled audio samples for use in semi-supervised training.  
    \item We empirically demonstrate the effectiveness of the proposed loss function to make quality predictions on unseen OOD speech synthesis systems in zero-shot, few-shot, and semi-supervised settings. 
\end{enumerate}

%% file: sections/method.tex
In this section, we present the proposed $\proposed$ criterion.
The method is motivated by the idea that the relative position of an audio sample in the ranking based on a partial list of training samples, which are ordered by their relative quality, is an important aspect of solving the MOS prediction task as opposed to only considering the absolute MOS value as in \cite{baseline}. 
Therefore, before delving into the specifics of the $\proposed$ criterion, the concept of relative position in the ranking and partial rank matrix needs to be explored in greater detail. 

\subsection{Relative position in the ranking and partial rank matrix}

Let us consider a list $\boldsymbol{l} = (l_1, l_2, l_3) = (1, 3, 2)$ where each element represents an absolute MOS value assigned to a different system. The list may not contain samples from all speech synthesis systems but a subset of them. Although the original ratings are ordinal values, we treat the MOS values as continuous for simplicity. To represent the relative position of each value with respect to all other values in the list, we define a matrix called the partial rank matrix. This matrix stores the position of each value in the list relative to every other value and also to itself. For example, 
the elements of the first row of the matrix are
$l_1 - l_1=0$, $l_1 - l_2=-2$, and, $l_1 - l_3=-1$, 
respectively. By extending this idea to all rows, we can construct the partial rank matrix for all values in $\boldsymbol{l}$, as shown in Equation \ref{eq:RL_matrix}. 

\begin{equation}
\centering
\label{eq:RL_matrix}
PR(\boldsymbol{l}) = 
 \begin{bmatrix} 
    0 & l_1 - l_2 & l_1 - l_3 \\
    l_2 - l_1 & 0 & l_2 - l_3 \\
    l_3 - l_1 & l_3 - l_2 & 0 \\
\end{bmatrix} 
=
\begin{bmatrix} 
    0 & -2 & -1 \\
    2 & 0 & 1 \\
    1 & -1 & 0 \\
\end{bmatrix} 
\end{equation}

A visual representation of the partial rank operation to the predicted and ground-truth MOS values is shown in Figure \ref{fig: Mos prediciton pipeline.}.

Matrix $PR(\boldsymbol{l})$ captures two fundamental pieces of information: directionality and magnitude. The sign in the matrix indicates the directionality, allowing us to determine whether the reference value is ranked higher or lower than all other indices. The magnitude simply represents the rank order difference, indicating {\em how much} higher or lower each value is than the reference value. 
Having established a solid foundation in understanding relative position in the rank, now we discuss the key aspects of the proposed $\proposed$ loss and its variants. 

\subsection{The $\proposed$ loss function}

During the training process, let us consider a batch of size $n$ containing $n$ input audio samples, denoted as $\boldsymbol{X} = (x_1, x_2, \ldots, x_n)$, and their corresponding MOS values. Our goal is to learn a prediction function that can closely estimate the MOS scores given the input audio signals. To achieve this, we assume the existence of a non-linear function $\Psi$ that approximates the MOS value on the basis of the provided audio, such that $\hat{y_i} = \Psi(x_i)$. We propose an objective function that minimizes the total losses with respect to the training data. The objective function is defined in Equation \ref{eq:formula}:
\begin{equation}
\label{eq:formula}
\centering
\mathcal{L}_{\proposed} = \left(\ \sum_{i=1}^n \sum_{j=1}^n \ \lambda * |PR_{ij}(\hat{\boldsymbol{Y}})\ - \ PR_{ij}(\boldsymbol{Y})|^{p}\ \right)^{1/p}
\end{equation}
\textbf{W}here: 
\begin{align}
\hat{\boldsymbol{Y}} &= (\hat{y}_1, \hat{y}_2, \ldots, \hat{y}_n),  \\
\hat{y}_i &= \Psi(x_i),  \\
\lambda &=
\begin{cases}
\label{eq:lambdac}
1 \quad \text{if } \{PR_{ij}(\hat{\boldsymbol{Y}}) \cdot PR_{ij}(\boldsymbol{Y})\} \leq 0, \\
\lambda_c <= 1 \quad \text{otherwise},
\end{cases} 
\end{align}
and $\lambda_c$ is a hyper-parameter. 
In Equation \ref{eq:formula}, the loss represents a measure of the difference between $ij$-th elements of the predicted $PR_{ij}(\hat{\boldsymbol{Y}})$ and the ground-truth matrices $PR_{ij}(\boldsymbol{Y})$, by utilizing a $p$-norm. 
Additionally, the weight factor $\lambda$ allows us to control the contribution of each index pair (i,j) in the total loss calculation. In one possible use case, if the two values ($PR_{ij}(\hat{\boldsymbol{Y}})$ and $PR_{ij}(\boldsymbol{Y})$) have the same sign (either both positive or both negative) in Eq.\ (\ref{eq:formula}), they are penalized less ($\lambda = \lambda_c < 1$) than in cases where they have opposite signs. In other words, if the MOS prediction model misclassifies the relative order of the $i$-th and $j$-th samples, we penalize more.

One benefit of using the loss in Eq.\ (\ref{eq:formula}) over the loss used by \cite{baseline} (which we call L1) is that the minimization takes into consideration other values and not just an individual value by incorporating the notion of relative positions in the ranking into the learning process. The model is explicitly encouraged to learn the correct rank order of the samples, whereas L1 regression does not consider the interaction between the samples. Furthermore, the number of comparisons (column) for each audio sample (row), in the $\proposed$ matrix is not restricted by the current batch size and can be easily extended by maintaining a cache of previous MOS values. We save the output of previous batches in a dictionary to be used for comparisons with the audio samples in the current batch. This is done because of the GPU memory limit. We call this variant ``Extended $\proposed$,-'' (E-$\proposed$ for short). 
The proposed $\proposed$ loss is a new approach for predicting the MOS of audio signals, is easy to implement, and can be used with any neural network architecture.
Lastly, we also investigate the combined E-$\proposed$ and L1 loss as shown in Eq.\ (\ref{eq:mixformula}). 
\begin{equation}
\centering
    \label{eq:mixformula}
      \mathcal{L} =\ \ \alpha * \mathcal{L}_{E-\proposed}\ +\ \beta * \left(\ \sum_{i=1}^n \ |\hat{y_i}\ - y_i|^{p}\ \right)^{1/p} \vspace{0.1cm}
\end{equation}

Similar to $\proposed$, $\max(0, |PR_{ij}(\hat{\boldsymbol{Y}})\ - \ PR_{ij}(\boldsymbol{Y})| - \gamma)$ is the loss used by the authors of UTMOS \cite{saeki22c_interspeech}. One major drawback is that their loss function does not always enforce correct rank order; i.e., even if the rank order is incorrect, the loss may be zero --  that is, all values less than $\gamma$ will be neglected.
In contrast, the $\proposed$ loss uses $\lambda_c$ to penalize less if the MOS prediction model orders the ranks of the $i$-th and $j$-th samples correctly. Lastly, if the MOS values of very similar systems are not reliable (assumption), then having a margin to ignore very small values is a good choice. This can be used as a regularizer in Equation \ref{eq:formula}.

\subsection{ \textbf{Pseudo MOS values selection algorithm for semi-supervised training}}

Assume we possess $n$ audio samples, each associated with their respective MOS values (labeled), and $m$ audio samples without the MOS values (unlabeled). In the semi-supervised setting, we initially train the model using the labeled samples through supervised learning. Subsequently, using the trained model, we estimate the MOS values for the unlabeled samples, which are referred to as pseudo MOS values. In the following phase, we merge the labeled and (selected) unlabeled samples and repeat the supervised learning as in the initial step. We iterate this procedure until a predefined stopping criterion is met. A straightforward selection algorithm would be to choose all unlabeled samples. One drawback is that not all the pseudo MOS values are accurate, which could destabilize the subsequent training phase. Therefore, a need arises for a better selection algorithm to pick pseudo MOS values that are likely to be correct. 

In this work, we propose a simple yet effective selection algorithm. Since it is challenging to define what is correct, we propose to simply balance the pseudo MOS values and call our method BAlanced pseudo MOS (BApMOS) selection. Given $m$ unlabeled audio samples and their corresponding pseudo MOS values $\hat{\boldsymbol{Y}} = (\hat{y}_1, \hat{y}_2, \ldots, \hat{y}_m)$, our method operates as follows: (i) We construct a histogram with $b$ bins (hyperparameter), each containing a count specified by $\boldsymbol{C} = (c_1, c_2, \ldots, c_b)$. If the resulting distribution is imbalanced, the method is prone to over-classify the majority group due to its higher prior probability. To address this issue, (ii) we randomly sample the minimum count, $\min(C)$, pseudo MOS values from each bin and discard the remaining values. The total number of selected pseudo MOS values for the iterative training is $b*\min(C)$. This simply ensures a balanced distribution of selected pseudo MOS values or uniform prior probability of the histogram.

\renewcommand{\arraystretch}{1}
\begin{table}[t]
\centering
\footnotesize
\caption{Summary of the different datasets used in this work.}
\label{tab:datastat}
\vspace{1mm}
\scalebox{0.97}{
\begin{tabular}{|c|c|c|c|c|c|}
\hline
{\multirow{2}{*}{\textbf{Dataset}}}  &  {\multirow{2}{*}{\textbf{Lang}}}  &\multicolumn{3}{c|}{\textbf{ \# Samples}} & \multirow{2}{*}{\parbox{1.4cm}{\centering\textbf{\footnotesize{\# ratings \\ per sample}}}} \\
&  & Train & Dev & Test &  \\ 
\hline
\multicolumn{6}{|c|}{\textbf{ Stage 1}} \\
\hline
BVCC \cite{baseline} & en & 4,974 & 1,066 & 1,066 & 8 \\
\hline

ASV2019 \cite{baseline} & en & - & - & 6,026 & 1-26 \\
\hline

BC2019 \cite{baseline} & ch & - & - & 450 & 10-17 \\
\hline

COM2018 \cite{baseline} & ja & - & - & 1,586 & 1-9 \\
\hline

\multicolumn{6}{|c|}{\textbf{ Stage 2}} \\
\hline
{\multirow{2}{*}{BC2019 \cite{huang22voicemos}}} & {\multirow{2}{*}{ch}} & \multirow{2}{*}{\parbox{1.6cm}{\centering Labeled: 136 \\ Unlabeled: 540}}
 & {\multirow{2}{*}{136}} & {\multirow{2}{*}{540}} & {\multirow{2}{*}{10-17}} \\

 &  &  &  &  &  \\
\hline

\cite{pinethesis} & Gitksan & - & - & 25 & 12 \\
\hline

\end{tabular}
}
\end{table}

%% file: sections/experimentationdetails.tex
\subsection{Experimental Design}

Two types of experiments are conducted in this paper: in Stage 1, an SSL model is first finetuned with the proposed criterion on the basis of the labeled training data. The evaluation is then performed on held-out data in the same domain as the training data. Zero-shot evaluations are also performed on three OOD sets that are not included in the training data.

In the Stage 2 experiments, we show and discuss the results of training a MOS prediction model with the proposed criterion on an OOD set, either by zero-shot, few-shot, or semi-supervised learning, and we also investigate the use of the BApMOS selection approach in the semi-supervised setting.

\subsection{Experimental Conditions}
\label{ssec:condition}

\renewcommand{\arraystretch}{1}
\begin{table*}[t]
\centering
\caption{Comparison of Stage 1 finetuned models, including prior work on the in-domain dataset.}
\label{tab:finetune}
\vspace{1mm}
\scalebox{0.85}{
\begin{tabular}{|p{5cm}|c|c|c|c|c|c|c|c|}
\hline
\multicolumn{1}{|c|}{\multirow{2}{*}{\textbf{Methods}}}  &  \multicolumn{4}{c|}{\textbf{ Utterance}} & \multicolumn{4}{c|}{\textbf{System}} \\ 

& MSE $\downarrow$ & LCC $\uparrow$ & SRCC $\uparrow$ & KTAU $\uparrow$ & MSE $\downarrow$ & LCC $\uparrow$ & SRCC $\uparrow$ & KTAU $\uparrow$ \\ 
\hline
L1 \cite{baseline} & \textbf{\phantom{0}0.227} & 0.868 & 0.866 & 0.690 & \textbf{\phantom{0}0.121} & 0.938 & 0.942 & 0.790 \\
\hline
UTMOS \cite{saeki22c_interspeech} & \phantom{0}5.870 & 0.869 & 0.866 & 0.687 & \phantom{0}4.810 & 0.948 & 0.951 & 0.806 \\
\hline
$\mathcal{L}_{\proposed}$, $\lambda_c=1.0$ & 10.670 & 0.879 & 0.878 & 0.704 & \phantom{0}8.800 & 0.951 & 0.951 & 0.811 \\
\hline
$\mathcal{L}_{E-\proposed}$, $\lambda_c=1.0$ & 12.320 & 0.881 & \textbf{0.881} & 0.707 & 10.120 & 0.947 & 0.949 & 0.805 \\
\hline
$\mathcal{L}_{E-\proposed}$, $\lambda_c=0.1$ & \phantom{0}7.240 & 0.872 & 0.869 & 0.692 & \phantom{0}6.260 & 0.944 & 0.941 & 0.800 \\
\hline
$\mathcal{L}_{E-\proposed}$, $\lambda_c=0.0$ & \phantom{0}3.490 & 0.602 & 0.862 & 0.684 & \phantom{0}2.320 & 0.643 & 0.920 & 0.760 \\
\hline
$\mathcal{L}$, $\lambda_c=1.0$, $\beta=0.01$ & \phantom{0}0.307 & \textbf{0.883} & \textbf{0.881} & \textbf{0.710} & \phantom{0}0.229 & \textbf{0.953} & \textbf{0.952} & \textbf{0.813} \\
\hline
$\mathcal{L}$, $\lambda_c=0.1$, $\beta=0.01$  & \phantom{0}0.490 & 0.874 & 0.871 & 0.700 & \phantom{0}0.490 & 0.937 & 0.938 & 0.790 \\
\hline
$\mathcal{L}$, $\lambda_c=0.0$, $\beta=0.01$  & \phantom{0}0.300 & 0.820 & 0.880 & 0.700 & \phantom{0}0.200 & 0.874 & 0.940 & 0.792 \\
\hline
\end{tabular}}
\end{table*}

\renewcommand{\arraystretch}{1}
\begin{table*}[t]
\centering
\caption{Comparison of zero-shot capabilities of $\proposed$ Stage 1 finetuned Wav2Vec2.0 model, its variants, and results from prior work on three out-of-domain datasets.}
\label{tab:finetune-ood}
\vspace{1mm}
\scalebox{0.85}{
\begin{tabular}{|p{5cm}|r|r|r|r|r|r|r|r|r|r|r|r|}
\hline

\multicolumn{1}{|c|}{\multirow{3}{*}{\textbf{Methods}}}  &  \multicolumn{12}{c|}{\textbf{Utterance}} \\ 
& \multicolumn{4}{c|}{\textbf{ASV2019}} & \multicolumn{4}{c|}{\textbf{BC2019}} & \multicolumn{4}{c|}{\textbf{COM2018}} \\
& MSE $\downarrow$ & LCC $\uparrow$ & SRCC $\uparrow$ & KTAU $\uparrow$ & MSE & LCC & SRCC & Ktau & MSE & LCC & SRCC & KTAU \\ 
\hline
L1 \cite{baseline} & \textbf{1.498} & \textbf{0.470} & 0.491 & 0.352 & 3.672 & 0.553 & 0.559 & 0.409 & 1.200 & 0.476 & 0.423 & 0.297 \\
\hline
UTMOS \cite{saeki22c_interspeech} & 4.610 & 0.462 & 0.479 & 0.342 & 26.990 & 0.658 & 0.684 & 0.489 & 14.750 & 0.463 & 0.431 & 0.307 \\
\hline
$\mathcal{L}_{\proposed}$, $\lambda_c=1.0$ & 8.430 & 0.464 & 0.475 & 0.339 & 45.250 & 0.649 & 0.681 & 0.493 & 25.520 & 0.466 & 0.436 & 0.309 \\
\hline

$\mathcal{L}_{E-\proposed}$, $\lambda_c=1.0$ & 9.010 & 0.464 & 0.479 & 0.342 & 51.800 & 0.635 & 0.654 & 0.464 & 29.090 & 0.502 & 0.463 & 0.331 \\
\hline
$\mathcal{L}_{E-\proposed}$, $\lambda_c=0.1$ & 2.750 & \textbf{0.470} & \textbf{0.499} & \textbf{0.357} & 19.510 & 0.637 & \textbf{0.686} & \textbf{0.500} & 6.34 & \textbf{0.515} & \textbf{0.490} & \textbf{0.350} \\
\hline
$\mathcal{L}_{E-\proposed}$, $\lambda_c=0.0$ & 4.500 & 0.253 & 0.480 & 0.342 & 38.100 & 0.604 & 0.651 & 0.467 & 2.64 & 0.401 & 0.443 & 0.315 \\
\hline

$\mathcal{L}$, $\lambda_c=1.0$, $\beta=0.01$ & 1.800 & 0.471 & 0.486 & 0.347 & \textbf{2.820} & 0.646 & 0.663 & 0.472 & \textbf{0.81} & 0.467 & 0.431 & 0.306 \\
\hline
$\mathcal{L}$, $\lambda_c=0.1$ , $\beta=0.01$ & 2.280 & 0.448 & 0.463 & 0.329 & 3.28 & 0.643 & 0.673 & 0.484 & 0.810 & 0.437 & 0.421 & 0.297 \\
\hline
$\mathcal{L}$, $\lambda_c=0.0$ , $\beta=0.01$ & 1.660 & 0.413 & 0.467 & 0.333 & 2.650 & \textbf{0.669} & 0.664 & 0.480 & 0.740 & 0.442 & 0.416 & 0.295 \\
\hline

\end{tabular}}
\end{table*}

\noindent \textbf{Pretrained Model}: Our approach utilizes the pretrained w2v\_small model \cite{wav2vec2}, which has 95 million parameters and generates 768-dimensional output embeddings from an input audio sample. This model was trained on the standard Librispeech dataset \cite{librispeech}, which comprises 960 hours of speech data.

\vspace{1mm}
\noindent \textbf{Loss Function}: We use our proposed $\proposed$ loss as described in Eq.\ (\ref{eq:mixformula}). We perform all the experiments with $p=1$ and squared $p=2$ norm. Similar to \cite{baseline}, we have found that the $p=1$ almost always gives slightly better results. Therefore, we only report results with $p=1$. Furthermore, the values of 
$\lambda_c$\ , $\alpha$ and, $\beta$ are set to $1.0$, $1.0$ and, $0.0$ in Eqs.\ (\ref{eq:lambdac}) and (\ref{eq:mixformula}) respectively, unless mentioned otherwise. Lastly, in the case of E-$\proposed$, the contribution of the extended columns to the loss is scaled by $1/10$.

\vspace{1mm}
\noindent \textbf{Finetuning} For Stage 1, we finetune the Wav2Vec2.0 model on the BVCC \cite{BVCC} training set using the $\proposed$ loss unless mentioned otherwise. Similarly to \cite{baseline}, we average the frame-level features of the last Wav2Vec2.0 layer and apply a linear regressor on top of it. The entire resulting model is then finetuned to solve the MOS prediction task using the BVCC training dataset. 

We also further finetune the Stage 1 model, best weights, on different OOD datasets for Stage 2 experiments. Three different sets of finetuning loss function configurations are used:  $\proposed$ / $\proposed$, L1 / L1 and, $\proposed$ / L1. 

Stage 2 finetuning consists of one of three setups: zero-shot, few-shot, or a semi-supervised scenario. In the zero-shot scenario, no finetuning is done i.e., 0 labeled and 0 unlabeled samples. In the few-shot scenario, small numbers of labeled samples are used for finetuning. In the semi-supervised setting, we generate predicted pseudo MOS values on the available unlabeled samples either using the Stage 1 or Stage 2 finetuned models. Then, we use these pseudo MOS values combined with the real scores to finetune the model further. During Stage 2 finetuning, we evaluate the model after each epoch, and if and only if the Spearman rank correlation coefficient (SRCC) metric improves on the development set, we regenerate the pseudo MOS values and continue finetuning.

\vspace{1mm}
\noindent \textbf{Dataset for Stage 1}: We evaluate the performance of our approach trained using the BVCC dataset, which was derived from a comprehensive listening test conducted by \cite{BVCC}. The dataset consists of 7,106 audio samples from 187 systems, including text-to-speech synthesis, voice conversion, and natural speech.  Each sample has eight ratings, which are averaged to obtain a MOS label for that sample. Listeners rated samples on a discrete scale from 1 (very bad) to 5 (very good) in terms of naturalness. We use the same training, development, and test sets as \cite{baseline}, preserving a distribution of $70\% /15\% /15\% (4,974/1,066/1,066)$. To assess the generalization ability of our approach, similar to \cite{baseline}, we also tested the BVCC-trained models on three OOD listening test datasets: ASV2019 \cite{asv2019} (English), BC2019 \cite{wu2019blizzard} (Mandarin Chinese), and COM2018 \cite{wang2018comparison} (Japanese). Testing was conducted in a zero-shot manner; i.e., without any further finetuning on these three OOD datasets. This evaluation protocol allows us to examine how well the model performs on unseen OOD data that is different from the training domain.

\vspace{1mm}
\noindent \textbf{Dataset for Stage 2}: For the Stage 2 finetuning experiments, we adopt the OOD track dataset from the Interspeech 2022 VoiceMOS challenge \cite{huang22voicemos}, which is the same original data as BC2019 except with different splits: there are 136 labeled training samples and 540 audio-only unlabeled training samples for use in semi-supervised training,  including an ``unlabeled training'' set. We also use a dataset from \cite{pinethesis}, consisting of five samples from each of four TTS systems and natural reference speech in the Gitksan language, an Indigenous language of Canada, for testing our approach on a real low-resource language.
Table \ref{tab:datastat} shows the statistics of all the datasets used in this work.

\vspace{1mm}
\noindent \textbf{Metrics}: Similar to \cite{baseline, huang22voicemos}, to evaluate MOS prediction models, we employ four widely used metrics: mean squared error (MSE), linear correlation coefficient (LCC), Spearman rank correlation coefficient (SRCC), and Kendall's Tau rank correlation (KTAU). 
The LCC, SRCC, and KTAU values range from -1 to 1, with values closer to 1 indicating a better correlation between predicted and ground-truth values. Among them, SRCC and KTAU are more useful metrics for our proposed loss function since MSE and LCC are dependent on absolute MOS values.

%% file: sections/results.tex
\renewcommand{\arraystretch}{1}
\begin{table*}[ht]
\centering
\caption{Testing the $\proposed$ method in zero-shot, few-shot and, semi-supervised settings on a dataset \cite{huang22voicemos}. E-$\proposed$ with $\lambda_c=0.1$ configuration is used for Stage 1 and Stage 2 finetuning. The results are averaged over three runs with random seeds. The row marked with * model is trained with the pseudo MOS values generated only once at the starting.}
\label{tab:semisupervised}
\vspace{1mm}
\scalebox{0.9}{
\begin{tabular}{|c|c|r|r|r|r|r|r|r|r|r|r|r|r|}
\hline

\multicolumn{1}{|c|}{\multirow{3}{*}{\parbox{1.7cm}{\centering\textbf{Number of \\ labeled \\ samples}}}}  & \multicolumn{1}{|c|}{\multirow{3}{*}{\parbox{1.7cm}{\centering\textbf{Number of \\ unlabeled \\ samples}}}}  & \multicolumn{12}{c|}{\textbf{1st finetuning loss / 2nd finetuning loss}} \\ 

& & \multicolumn{4}{c|}{$\proposed$ / $\proposed$} & \multicolumn{4}{c|}{L1 / L1} & \multicolumn{4}{c|}{$\proposed$ / L1} \\

& & MSE $\downarrow$ & LCC $\uparrow$ & SRCC $\uparrow$ & KTAU $\uparrow$ & MSE & LCC & SRCC & KTAU & MSE & LCC & SRCC & KTAU \\ 
\hline
\multicolumn{14}{|c|}{\textbf{Zero-shot setting}} \\
\hline
0 & 0 & 16.350 & 0.617 & \textbf{0.651} & 0.457 & 3.150 & 0.532 & 0.538 & 0.387 & 16.350 & 0.617 & \textbf{0.651} & 0.457  \\
\hline
\multicolumn{14}{|c|}{\textbf{Few-shot setting}} \\
\hline
10 & 0 & 13.160 & 0.657 & 0.690 & 0.486 & 0.980 & 0.715 & 0.708 & 0.509 & 0.640 & 0.701 & \textbf{0.744} & 0.542 \\
\hline
136 & 0 & 6.960 & 0.873 & \textbf{0.842} & 0.652 & 0.660 & 0.845 & 0.825 & 0.632 & 0.750 & 0.865 & \textbf{0.843} & 0.652 \\
\hline
\multicolumn{14}{|c|}{\textbf{Semi-supervised setting}} \\

\hline
0* & 136* & 12.414 & 0.651 & 0.686 & 0.484 & - & - & - & - & - & - & - & - \\
\hline
0 & 136 & 4.000 & 0.807 & \textbf{0.778} & 0.580 & 13.050 & 0.721 & 0.744 & 0.550 & 9.910 & 0.720 & 0.773 & 0.572 \\
\hline
0 & 676 & 1.980 & 0.768 & \textbf{0.778} & 0.582  & 11.190 & 0.701 & 0.747 & 0.551 & 23.920 & 0.623 & 0.751 & 0.553 \\
\hline

10 & 126 & 0.750 & 0.783 & \textbf{0.786} & 0.582 & 2.750 & 0.703 & 0.686 & 0.493 & 2.900 & 0.675 & 0.705 & 0.509 \\
\hline
10 & 666 & 1.160 & 0.770 & \textbf{0.782} & 0.583 & 8.790 & 0.663 & 0.696 & 0.503 & 11.910 & 0.606 & 0.672 & 0.483 \\
\hline
136 & 540 & 0.650 & 0.858 & \textbf{0.839} & 0.646 & 0.660 & 0.845 & 0.825 & 0.632 & 1.330 & 0.860 & \textbf{0.840} & 0.650\\
\hline
\end{tabular}}
\end{table*}

\renewcommand{\arraystretch}{1}
\begin{table*}[ht]
\centering
\caption{Testing the $\proposed$ method on Gitksan language \cite{pinethesis}, similar to Table \ref{tab:semisupervised}. Readers must keep in mind that because only 25 samples were available, we discard the MOS values and treat them as unlabeled samples in the semi-supervised setting. However, for development and testing purposes, we use the ground-truth MOS values for comparison.}
\label{tab:semisupervised-lr}
\vspace{1mm}
\scalebox{0.9}{
\begin{tabular}{|c|c|r|r|r|r|r|r|r|r|r|r|r|r|}
\hline

\multicolumn{1}{|c|}{\multirow{3}{*}{\parbox{1.7cm}{\centering\textbf{Number of \\ labeled \\ samples}}}}  & \multicolumn{1}{|c|}{\multirow{3}{*}{\parbox{1.7cm}{\centering\textbf{Number of \\ unlabeled \\ samples}}}}  & \multicolumn{12}{c|}{\textbf{1st finetuning loss / 2nd finetuning loss}} \\ 

& & \multicolumn{4}{c|}{$\proposed$ / $\proposed$} & \multicolumn{4}{c|}{L1 / L1} & \multicolumn{4}{c|}{$\proposed$ / L1} \\

& & MSE $\downarrow$ & LCC $\uparrow$ & SRCC $\uparrow$ & KTAU $\uparrow$ & MSE & LCC & SRCC & KTAU & MSE & LCC & SRCC & KTAU \\ 
\hline
\multicolumn{14}{|c|}{\textbf{Zero-shot setting}} \\
\hline
0 & 0 &  6.210 & 0.810 & 0.790 & 0.640 & 0.940 & 0.760 & 0.690 & 0.530 & 6.210 & 0.810 & 0.790 & 0.640\\
\hline
\multicolumn{14}{|c|}{\textbf{Semi-supervised setting}} \\
\hline
0 & 25 & 5.440 & 0.835 & 0.851 & 0.696 & 4.400 & 0.717 & 0.763 & 0.608 & 1.200 & 0.791 & 0.848 & 0.680  \\
\hline

\end{tabular}}
\end{table*}

\subsection{Stage 1 experiment: in-domain vs.\ out-of-domain}

Table \ref{tab:finetune} shows comparison results of the Stage 1 models on the in-domain BVCC test dataset. First, we can confirm that since the predictive models using the proposed $\proposed$ loss and its variant ($\mathcal{L}_{\proposed}$ and $\mathcal{L}_{E-\proposed}$) do not take into account the absolute MOS values during the learning process, they naturally result in larger MSEs, but this outcome is expected. Next, comparing the values of LCC, KTAU, and SRCC, we can confirm that the correlation coefficients of the proposed methods ($\mathcal{L}_{\proposed}$,  $\mathcal{L}_{E-\proposed}$, and $\mathcal{L}$) are comparable to or even slightly higher than those of L1 and UTMOS when appropriate $\lambda_c$ values are utilized. Finally, the results of using a loss $\mathcal{L}$ that also takes L1 into account at the same time naturally confirms that the MSE is also reduced. In summary, if one wants to know only the rank ordering, the proposed loss function is sufficient; if one wants to approximate the MOS values as well, L1 is necessary.

Table \ref{tab:finetune-ood} shows zero-shot comparison results of the Stage 1 models on the three OOD test datasets. First, this evaluation is done in a zero-shot manner, so naturally, the overall correlation coefficients are lower, and the MSEs are larger. We then see that the models trained with the proposed loss function have a similar level of correlation coefficients to the case trained with L1 evaluated on the OOD test sets. Some minor but consistent improvement is also observed. For instance, a system using the $\mathcal{L}_{E-\proposed}$ with $\lambda_c=0.1$ has consistently better rank correlations (SRCC and KTAU) than L1 and UTMOS on three out of three OOD datasets. The improvement is more evident in unseen languages, that is, BC2019 and COM2018. 

Finally, regarding the combined $\proposed$ and L1 loss, we see a small amount of degradation concerning the rank correlations. This suggests that the two losses are not working in tandem and that minimizing the absolute values is not a good strategy for solving the MOS prediction task in the OOD setting. To summarize, E-$\proposed$ with $\lambda_c=0.1$ has the best generalization ability given its performance gains on unseen languages.

\subsection{Stage 2 experiment: a comparison of zero-shot, few-shot, and semi-supervised settings}

In the Stage 2 experiment, we analyze the performance of MOS predictors in zero-shot, few-shot, and semi-supervised settings. 
As explained in Section \ref{ssec:condition}, we finetune the Stage 1 model using small amounts of labeled samples for the few-shot setting, whereas we generate pseudo MOS labels for unlabeled training audio samples and finetune a model by mixing the labeled samples and pseudo labeled ones for the semi-supervised setting.  

Table \ref{tab:semisupervised} shows results on the BC2019 dataset \cite{huang22voicemos}. First, we see that both few-shot and semi-supervised learning improved correlation coefficients. This is true even for semi-supervised cases where no labeled samples are used.  
As for the combinations of the losses used for the first and second finetuning, we see that the models using the $\proposed$ loss for the first finetuning generally resulted in higher rank correlation coefficients after the second finetuning. This trend can be clearly seen from the SRCC values in the table. This demonstrates the generalization ability of the $\proposed$ loss. Interestingly, the semi-supervised setting with the $\proposed$ / $\proposed$ condition has smaller MSE values as well. 
The next observation is that increasing the unlabeled data for semi-supervised learning (136 to 676 samples and 126 to 666 samples) does not result in any performance gains. This could be attributed to using all unlabeled samples with their pseudo MOS values during finetuning. Lastly, the empirical results show that iteratively regenerating the pseudo MOS values is necessary and is more accurate than if the pseudo MOS values are generated only once at the starting as shown in Table \ref{tab:semisupervised}, in the row marked with *.

Semi-supervised learning is particularly helpful for low-resource language scenarios since it is not straightforward to find native listeners. We therefore additionally analyzed the performance of MOS predictors in zero-shot and semi-supervised settings on a MOS dataset in the Gitksan language \cite{pinethesis}. Table \ref{tab:semisupervised-lr} shows results of the zero-shot and semi-supervised inference on the MOS dataset for the Gitksan language. We can see the same trend – the semi-supervised learning without using any labeled samples improved the prediction performance, and the $\proposed$ / $\proposed$ condition resulted in the highest rank correlation coefficients.

\subsection{Stage 2 experiment: a comparison of semi-supervised learning using the BApMOS selection strategy}

Next, we compare semi-supervised learning with and without the proposed BApMOS selection algorithm on the BC2019 dataset as shown in Table \ref{tab:bapmos}. We only report the SRCC values. 

First, by comparing the results of semi-supervised learning on 136 samples with and without the BApMOS selection algorithm, we can see that the proposed BApMOS selection algorithm works effectively. It considerably boosts the performance over simply using all of the pseudo labels. As expected, it is not as good as few-shot learning which uses the ground-truth labels. Furthermore, increasing the number of unlabeled samples from 136 to 676 results in a slight performance gain, which was not the case earlier. This again proves the importance of a selection algorithm rather than just using all the unlabeled samples.

Furthermore, we make two observations: (i) diversity of selected pseudo-MOS values is detrimental to the performance of the $\proposed$ method. When using $676$ unlabeled samples, increasing the number of bins boosts the performance significantly. (ii) The total number of selected samples is more important than diversity if there are very few selected samples, as in the case of $30$ and $20$ bins in $676$ and $136$ unlabeled samples, respectively. 
Since this is a promising result, we hope that using better selection methods will result in additional performance gains, as shown by the success of semi-supervised learning methods in the past \cite{de2016highconfidence}. We leave this for future work.

\renewcommand{\arraystretch}{1}
\begin{table}[t]
\centering
\caption{Testing the BApMOS selection algorithm for $\proposed/\proposed$ configurations, similar to Table \ref{tab:semisupervised}. Here the SRCC metric was used to compare the performance. }
\label{tab:bapmos}
\vspace{1mm}
\scalebox{0.93}{
\begin{tabular}{|c|c|c|c|c|c|c|}
\hline
\multicolumn{1}{|c|}{\multirow{3}{*}{\parbox{1.3cm}{\centering\textbf{\small Number of \\ labeled \\ samples}}}}  & \multicolumn{1}{|c|}{\multirow{3}{*}{\parbox{1.3cm}{\centering\textbf{\small Number of \\ unlabeled \\ samples}}}} &  & \multicolumn{4}{c|}{\multirow{1}{*}{\parbox{4.0cm}{\centering\textbf{\small Number of bins for a histogram }}}} \\

&&&&&& \\
& & \_ & 5 & 10 & 20 & 30 \\ 
\hline
\multicolumn{7}{|c|}{\textbf{\small Few-shot setting}} \\
\hline
136 & 0 & \textbf{0.842} & - & - & - & - \\
\hline
\multicolumn{7}{|c|}{\textbf{\small Semi-supervised setting}} \\
\hline
0 & 136 & 0.778 & - & - & - & -\\
\hline
0 & 676 & 0.778 & - & - & - & - \\
\hline
\multicolumn{7}{|c|}{\textbf{\small Semi-supervised setting + BApMOS selection}} \\
\hline
0 & 136 &  & \textbf{0.804} & \textbf{0.800} & 0.800 & -\\
\hline
0 & 676 &  & 0.780 & 0.797 & \textbf{0.809} & 0.799\\
\hline

\end{tabular}}
\end{table}

%% file: sections/discussion.tex
The MOS test is affected by not only the quality of the speech, but also by the various contexts during the listening test, which cause MOS values to fluctuate. The need to model the influence of this context is an important decision regarding automatic MOS prediction.

If we believe that the variation in MOS values also needs to be modeled in the current target context, then we will need to use the MOS values as supervised labels for training. However, since this policy learns a context-dependent model, it is not expected to generalize to test sets in different contexts.

On the other hand, we found that the MOS prediction model generalizes better to the OOD test set when the learning criterion is based on the rank order of the systems, rather than using context-dependent MOS values directly as the learning target. Although the context of that OOD test set cannot be properly considered in a zero-shot manner, we show that some context information can be captured by semi-supervised learning if unlabeled speech data is available.

The semi-supervised learning proposed in this paper has room for improvement. Specifically, the proposed semi-supervised learning used unlabeled speech data for training, but the development set still contains labeled speech. By using unlabeled speech data even in the development set, the semi-supervised learning of the MOS prediction model will be more useful. Lastly, as of now, we have not selected the samples based on any criterion other than simply making the prior of the histogram uniform. We randomly select the samples from the bin, but if instead a heuristic is used, that could lead to further improvements. One possible heuristic is to drop any sample whose relative pseudo MOS value is higher than the natural speech. There could be more ways to do selection but we leave this for future work.

%% file: sections/appendix.tex
\section{Stability} 

\begin{figure*}[ht]
\centering
\includegraphics[width=1\columnwidth]{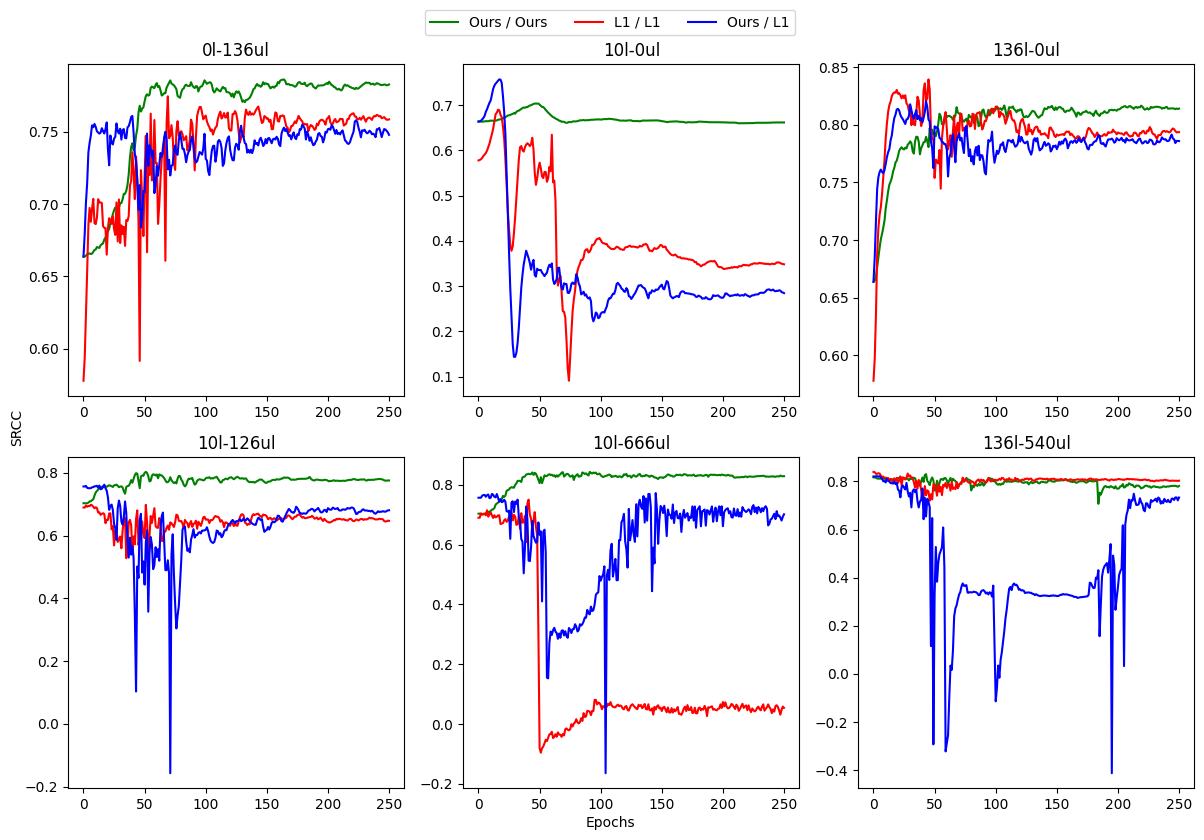}
\caption{Few-shot and semi-supervised SRCC trend on the BC2019 validation set. Better viewed in color.}
\label{fig: Semi supervised SRCC results.}
\end{figure*}

Figure \ref{fig: Semi supervised SRCC results.} plots the epochs vs.\ the SRCC of the validation set used in the Stage 2 finetuning. There is a clear pattern of more stable SRCC values in the $\proposed$ / $\proposed$ setting than in the other two. This shows that both the pseudo MOS values and Stage 2 finetuning are stable and no severe overfitting takes place when the loss used is $\proposed$. This again demonstrates the generalization ability of $\proposed$ loss.

\renewcommand{\arraystretch}{1.5}
\begin{table*}[ht]
\centering
\caption{Wav2Vec2.0 models is used as a feature extractor with different loss functions.}
\label{tab:feature_extractor}
\scalebox{1}{
\begin{tabular}{|c|c|c|c|c|c|c|c|c|c|}
\hline

\multicolumn{1}{|c|}{\multirow{2}{*}{\textbf{Method}}}  & \multicolumn{1}{c|}{\multirow{2}{*}{\textbf{Loss}}}  & \multicolumn{4}{c|}{\textbf{ Utterance}} & \multicolumn{4}{c|}{\textbf{System}} \\ 

& & MSE $\downarrow$ & LCC $\uparrow$ & SRCC $\uparrow$ & Ktau $\uparrow$ & MSE & LCC & SRCC & Ktau \\ 
\hline 
\multicolumn{1}{|c|}{\multirow{3}{*}{Nonlinear}} & L1 \cite{baseline} & 0.40 & 0.75 & 0.75 & 0.56 & 0.18 & 0.86 & 0.87 & 0.68 \\
\cline{2-10}
& UTMOS \cite{saeki22c_interspeech} & 10.05 & 0.792 & 0.794 & 0.606 & 9.24 & 0.906 & 0.907 & 0.732 \\
\cline{2-10}
& $\proposed$ & 9.86 & 0.799 & \textbf{0.798} & 0.611 & 8.94 & 0.928 & \textbf{0.929} & 0.770 \\
\hline
\multicolumn{1}{|c|}{\multirow{3}{*}{Linear}} & L1 \cite{baseline} & 0.38 & 0.82 & \textbf{0.82} & 0.63 & 0.15 & 0.90 & 0.89 & 0.725 \\
\cline{2-10}
& UTMOS \cite{saeki22c_interspeech} & 9.50 & 0.795 & 0.799 & 0.611 & 9.20 & 0.906 & 0.906 & 0.733 \\
\cline{2-10}
& $\proposed$ & 9.60 & 0.81 & 0.81 & 0.62 & 9.30 & 0.92 & \textbf{0.91} & 0.74 \\
\hline

\end{tabular}}
\end{table*}

\section{W2v model used as a feature extractor}
We utilize the w2v SSL model to extract features from input audio sample. First, the raw waveform of an audio is fed into the SSL model to obtain frame level features. We follow the SUPERB benchmark settings \cite{yang2021superb}, where features from each layer are linearly weighted and averaged to obtain the output features. Similarly to \cite{baseline}, we average the frame-level output features and apply one of two (i)linear or (ii)non-linear layers to solve the MOS values prediction task. The BVCC dataset is used in training the prediction layer and testing. The results show that our proposed $\proposed$ loss either outperforms or is comparable to the L1 loss \cite{baseline} and always outperforms the UTMOS \cite{saeki22c_interspeech} loss, as shown in Table \ref{tab:feature_extractor}. Notably, the performance improvements are significant in the non-linear case, which shows that features learned using the $\proposed$ loss generalize better compared to the L1 loss. Similarly, UTMOS also shows improvement over L1 although less so than $\proposed$, which again demonstrates the generalization capability of losses using relative location in general. The results of the non-linear case are more important because finetuning the model makes the learnable function non-linear. We also test w2v model, from DDOS \cite{tseng22b_interspeech}, which was additionally pre-trained on the TTS generated audio samples (DAPT). The SRCC values were considerably worse compared to using the original w2v model without DAPT. We are not sure of the reason and leave it to the future work.

\section{Should the research community treat MSE and LCC as reliable evaluation metrics?}

Assume a scenario where we keep the relative order of audio samples, but shift the MOS values by a constant amount (100). If we re-evaluate using the MOS prediction system again, the mean squared error (MSE) would exhibit a substantial increase, while the SRCC would remain unaffected. This suggests that accurately predicting the relative naturalness order of synthesis systems is of greater significance than determining their absolute MOS values. 

In all of our experiments we observe that the MSE value (lower is better) of the $\proposed$ (or UTMOS) method is always very high (10-20 times compared to the L1 loss). However, $\proposed$ is always better at predicting the monotonic relationship (SRCC) between the predicted and ground truth values as shown in Table \ref{tab:feature_extractor}, \ref{tab:finetune}, \ref{tab:finetune-ood} and, \ref{tab:semisupervised}. Furthermore, setting the value of $\lambda=0.0$ results in lower LCC as shown in Table \ref{tab:finetune} and \ref{tab:finetune-ood} because LCC also takes into account the absolute values during the calculation. These two observations demonstrate that MSE and LCC are not reliable metrics to compare different MOS prediction systems.